# *Chemically Resolved Electrical Measurements using X-ray Photoelectron Spectroscopy*

**Hagai Cohen**

Chemical Research Support

The Weizmann Insitute of Science, Rehovot 76100, Israel

## Introduction

A novel approach is proposed, where energy filtered electrons, carrying both chemical identity and electrical information, serve as fine and flexible electrodes in direct electrical measurements. The method, termed 'chemically resolved electrical measurements' (CREM), is simple and general, demonstrated here with a slightly modified X-ray photoelectron spectrometer. Selected *sub-surface* regions are electrically analyzed and I-V curves of self-assembled monolayers, free of substrate and top contact contributions, are derived with no need for improved metallic substrates. Unique electrical information is available with this method, further supported by powerful in-situ analytical capabilities and improved top contact performance.

----------------------

Shrinking down device dimensions[1], for example in molecular electronics[2-7], requires breakthrough concepts in electrical testing and characterization: The electrode size necessarily reaches a principal limit at the nanometer scale. Moreover, while multi-component designs of the single chip become broadly used, the existing tools for direct electrical measurements are generally limited in resolving *internal* regions[8].

Electrical characterization of molecular systems is of particular difficulty. Even 'simple' mono-molecular layers, proposing well established binding options to 'external world' contacts, appear to be tricky objects for conductance measurements[3,4]. Certainly, when non-metallic substrates are chemically preferable, the electrical characterization becomes even harder. This problem reflects a general aspect of the electrical measurement - its typically high sensitivity to structural imperfections and contact details[1]. An increasing need in electrical probes that incorporate structural and analytical capabilities is therefore raised by the recent attempts to develop templates for molecular electronics.

A number of works[9-12], in particular the surface charge spectroscopy (SCS) by W.M. Lau and coworkers, have already shown that the line-shifts occurring upon charging in X-ray photoelectron spectroscopy (XPS) can be used to extract position of atoms. In a recent work[13,14], based on controlled surface charging (CSC) in XPS, nm resolution depth profiling, as well as fine site differentiation in the lateral dimensions, were demonstrated. Electrical properties were also investigated [15-17] by SCS in XPS. However, those works remained limited in providing quantitative electrical analysis as long as ohmic effects were neglected.

The present work proposes a more consistent and comprehensive framework for electrical studies based on CSC. XPS line-shifts are correlated here with the current detected on the back contact of the sample, while, conceptually, the circuitry is generalized to include non-wired currents (electron beams). The additional input obtained by detecting the sample current, a technically trivial step, is found to significantly improve the control of the charging source and, in fact, provide a key piece of information for deeper understanding of the experiment. Chemically resolved electrical measurements (CREM), i.e. element specific I-V curves, are derivable, proposing diverse and unique applications. In the following, the main idea and its realization are explained. The net I-V curve of a thin molecular layer is extracted, demonstrating some of the advantages of the present approach.

The electrical circuit here consists of two non-contact, surface oriented components plus a back contact to an ampermeter. Fig. 1a presents the main functions in the actual system. A simplified analogue circuit is given in Fig. 1b. The most unique component of the circuit is its multi-channel voltmeter; an electron spectrometer, which reads local potential values from line positions[13]. Different lines recorded by this spectrometer represent different chemical entities. Thus, with predetermined chemical characteristics, an inherent aspect of system design, each spectral line can serve as a channel for local potential measurements. This voltmeter can probe quite deep regions, down to ~15 nm[18]. Moreover, the assumptions on the 'already known' structure can be checked here, in-situ, by a powerful analytical technique, XPS.

The power supply in this circuit, an electron flood gun (eFG), is a distant source of soft electrons[19]. Though not physically attached to the specimen, it manifests an optimal electrical contact, free of defects and inter-atomic (condensed matter) interactions. When the effective circuit resistance is sufficiently high, the (finite) eFG source becomes voltage limited and hence functions as a battery[20]. At the limit of low resistance it is current limited, serving as a current supply. Note that one can always match the voltage-limited situation by adding back resistors. Having direct feedback on the actual surface potential, a full control of the effective eFG impact is established.

Measurements were performed on a slightly modified Kratos AXIS-HS setup. The eFG was controlled via filament current and two bias voltages, its beam diameter, on a scale of a few mm, subjected to variations upon changes in bias values. (A magnetic lens is used in this instrument; its effect on the eFG operation will be discussed elsewhere). An external Keithley 487 electrometer and additional known resistors were connected to the back contact of the sample. Spectral line shifts were derived at accuracy of 10-50 meV, depending on noise level and spectral stability[21]. Conductive double-sided carbon tape, and in several cases InGa, were used at the sample-holder back contact (no top contacts were used here). Studying relatively large and high quality homogeneous samples, lateral effects could be largely suppressed (as verified from test experiments near sample edges, not shown). The procedures for self-assembly of the high-quality molecular layers on p-Si, here eicosenyl amino silane ($NH_2C_{20}Si$), have been described elsewhere[22]. Photovoltaic effects were also accounted for (not shown here) by varying the X-ray flux and by illuminating the sample with an additional source of light. Their impact on the systems shown below was found to be negligible.

Circuitry performance is demonstrated by several I-V curves, Fig. 2. The surface potential at the Si wafer is given by the kinetic energy of the $Si(2p_{3/2})$ (non-oxidized) line. It is presented here on a relative scale with no attempt to determine the zero potential point. A good ohmic behavior is established by the highly doped n-Si wafer, curves a-d: The slope of the various curves is just the effective impedance between the sample surface and ground, practically independent of the choice of eFG parameters. Curve e presents a different case, recorded with p-Si(100). Exponential I-V dependence is observed here (see inset), arising from the back contact, where the double-sided tape creates a Schottky-type junction with the wafer. Both samples indicate regular circuitry performance, which obeys common rules and hence applicable to various systems.

An important feature, the elimination of top-contact problems in *direct* electrical measurements, is demonstrated in Fig.2. Yet, the main advantage of CREM arises from its capability to chemically discriminate the I-V information. In Fig. 3, four I-V curves represent three different regions at a Si surface: the wafer itself (Si), its surface oxide[23] (Si(ox) and O) and a carbonic overlayer[24] (C). Such curves can be easily translated into zone-specific data. For example, the resistance of the buried oxide layer is derived from a difference curve, Si(ox)-Si, yielding 0.83 MΩ/nm. Similarly, the carbonic layer resistance is found to be 0.53 MΩ/nm.

The accuracy of local potential determination is subjected to the actual potential gradients across the probed volume. Practically, however, this broadening effect is largely suppressed by the attenuation effects and further by analyzing *relative shifts* only, more specifically the shifts of the right hand-side (high kinetic energy) line-edge[13,25]. In the present case, the derived values can be assigned to the top of each corresponding layer. The accuracy of this assignment has been estimated with top markers to be better than 0.4 nm. Further improvement is possible by convoluting the bare line-shape with a function describing the potential drop and the corresponding signal attenuation[13,25].

To exemplify method advantages, consider measuring the conductance through a molecular layer, an intriguing subject, which exhibits major technical difficulties. Here, even with *non-metallic* substrates, all contact contributions are removable, as shown above. However, since molecular systems frequently present poor electrical stability, two drawbacks of the CREM technique arise: First, measurements are typically slow (a technical point to be highly improved in the future); second, the X-irradiation tends to induce sample damage, to which electric measurements are usually highly sensitive.

An attempt to overcome these difficulties is presented in Fig. 4, measuring a high-quality self-assembled monolayer (SAM)**,** $NH_2C_{20}Si$, on p-Si. The elements in Fig. 4a represent regions similar to those in Fig. 3. Energies are referred to the non-charged state (minimal X-ray power and eFG-off conditions), thus presenting an absolute potential scale. The 'non-damaged' limit has been achieved by cooling to ca 200 K and recording rapidly the line shifts (at a compromised accuracy)[27]. Nitrogen detection required slower scans, therefore not included in Fig. 4.

Fig. 4b shows difference I-V curves, 'C-Ox' for the 'contact free' hydrocarbon layer and 'Ox-Si' representing the buried oxide layer net response. The former curve is roughly exponential, suggesting that tunneling mechanisms take place here[6,26]. It is stressed that the given voltage values are those actually developed at the film surface. The current, on the other hand, includes also ballistic electrons, crossing the layer without reaching thermodynamic equilibrium. In the present measurement, remaining in the 'voltage limited' situation, the ballistic current is constrained to be roughly constant, on the order of a few nA only. Thus, the curve in Fig. 4b represents reliably the current associated with charges accumulated at the surface. A systematic investigation of conduction mechanisms, distinguishing for example the ballistic flow from slower processes, is an interesting problem for future studies based on this technique.

The present method capabilities largely depend on the chemical contrast within the sample structure, demonstrated here with nanometric layers. In principle, however, spatial selectivity of CREM may be 'pushed' even further, merging with the traditional question of atomic chemical shifts in XPS. Organic molecular structures, for example, frequently offer an extremely flexible chemistry, which might be exploited for fine electrical studies at the sub-molecular scale[28]. This challenging subject will be available when technical aspects of the measurement are improved.

Unique advantages of CREM are also expected in studies of transport mechanisms through thin dielectric spacers, as well as in studies of traps and space-charge distribution within semiconducting systems. It should be noted that the method is not limited to XPS-based set-ups. Additional spectroscopies of charged particles, where chemical and electrical information can be correlated, e.g. Auger electron spectroscopy (AES), can similarly become successful templates for CREM. Based on a general concept and conventional equipment, the proposed technique is believed to successfully replace existing electrical tools in various applications.

**Acknowledgements**

The author thanks J. Sagiv and R. Maoz for providing the high quality self-assembled monolayers. S. Reich is acknowledged for helpful discussions.

**Figure Captions**

Fig. 1: **The generalized electric circuitry:** A schematic illustration of: (a) the actual system; (b) the analog circuit. Dashed lines stand for non-wired currents. The power (the battery in 1b, including its internal resistor $r_{in}$) is supplied via flood of soft electrons (eFG in 1a), its energy and flux externally controllable. A distant multi-channel voltmeter (V in 1b) is established by analyzing (e-analyzer in 1a) internal electrons that are 'kicked-out' by means of an additional source of energy, here a monochromatic X-ray beam and an electronic lens system (the latter not shown). The ampermeter (A) and the external resistor ($R_{ext}$) are connected to the back contact of the sample. Additional components can be incorporated to the circuit at will.

Fig 2: **XPS-based electric measurements:** Trivial I-V curves, plotting the (negative) sample-current ($I_s$) vs kinetic energy (Ek) of the Si ($2p_{3/2}$) (non-oxidized) line: Curves a-d correspond to highly doped n-Si and curve e to p-Si. Back impedance is 1MΩ in all curves except for d, where $R_{ext}$=10MΩ. The varied eFG parameters are: (a) filament current; (b) bias voltages; (c) bias voltages in a different combination; (d) same as c, but with different back resistance; (e) filament current. The inset shows curve e on a log scale.

Fig. 3: **Chemically resolved I-V plots:** Electrical analysis across a Si-wafer surface, where each XPS line represents a selected region (see text). An arbitrary zero is chosen for the kinetic energy scale. Bias eFG is varied here in the range of 3.3-4.5 V.

Fig. 4: **Self-assembled layer conductance:** Application to $NH_2C_{20}Si$ self-assembled monolayer on p-Si, varying eFG voltages in the 2-5 V range: (a) Element specific plots representing selected regions: the organic layer (C), the Si-oxide (O) and the wafer (Si); (b) Difference curves, C-Ox and Ox-Si, exhibiting the net response of the organic monolayer and the (inner) oxide, respectively. The slight hysteresis in 4a arises from scanning a bit faster than the response time of the system (irreversible sample degradation is observable at longer time scales). Current evolution with time due to increasing density of defects has been evaluated and accounted for in 4b.

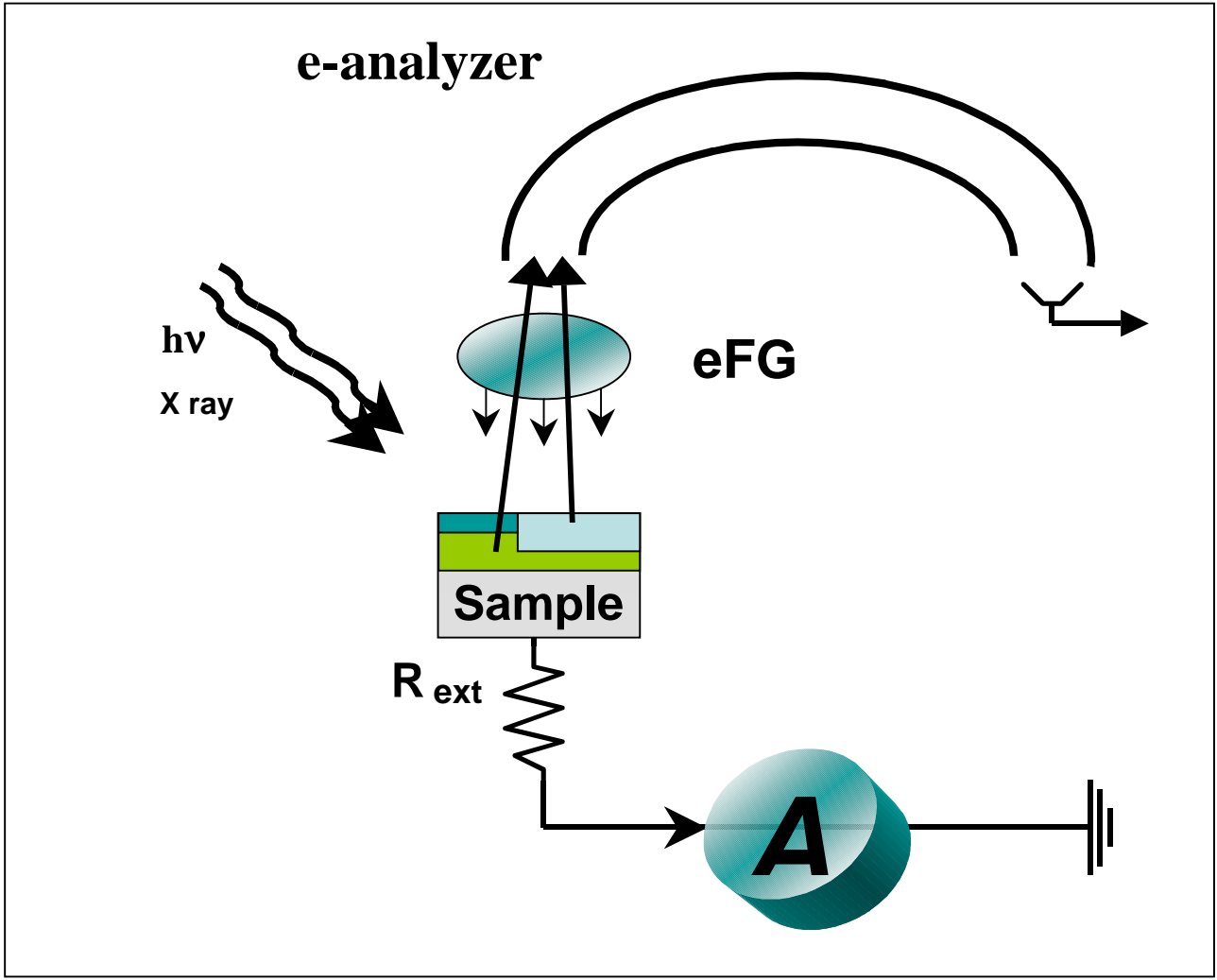

Fig. 1a

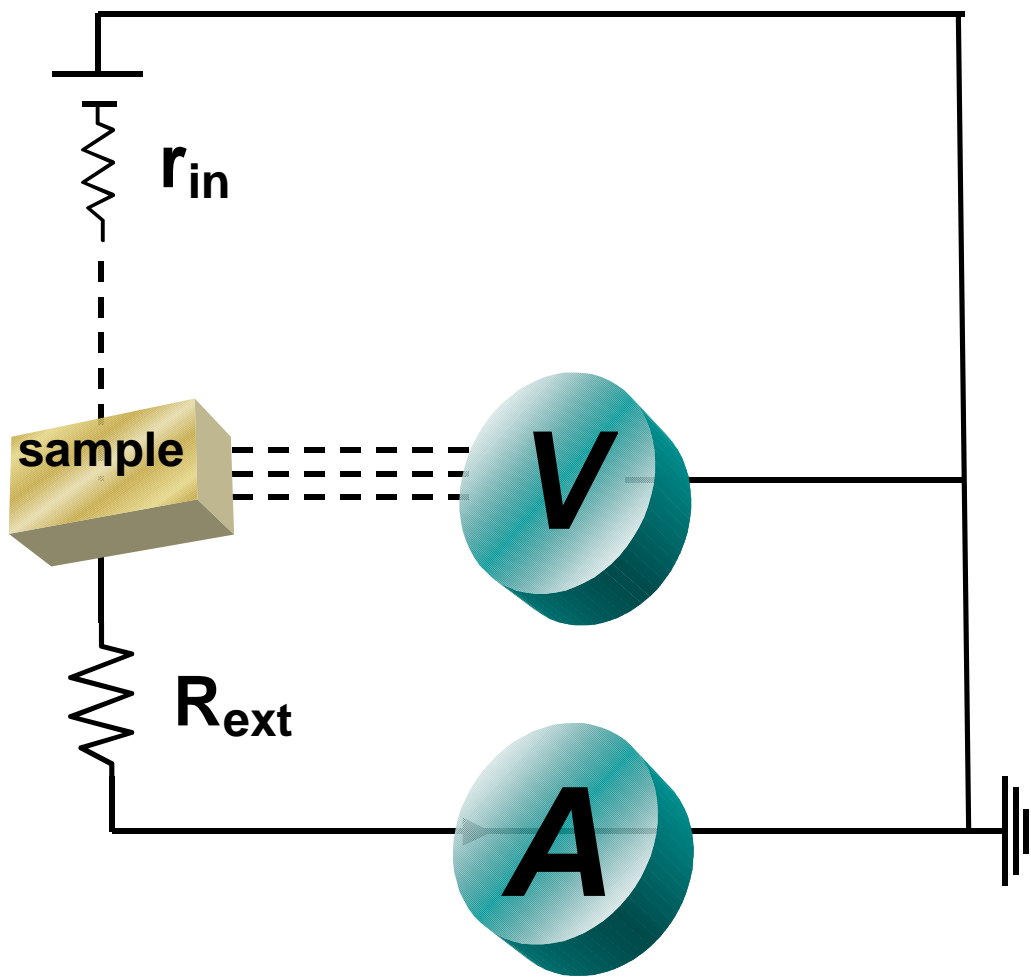

Fig. 1b

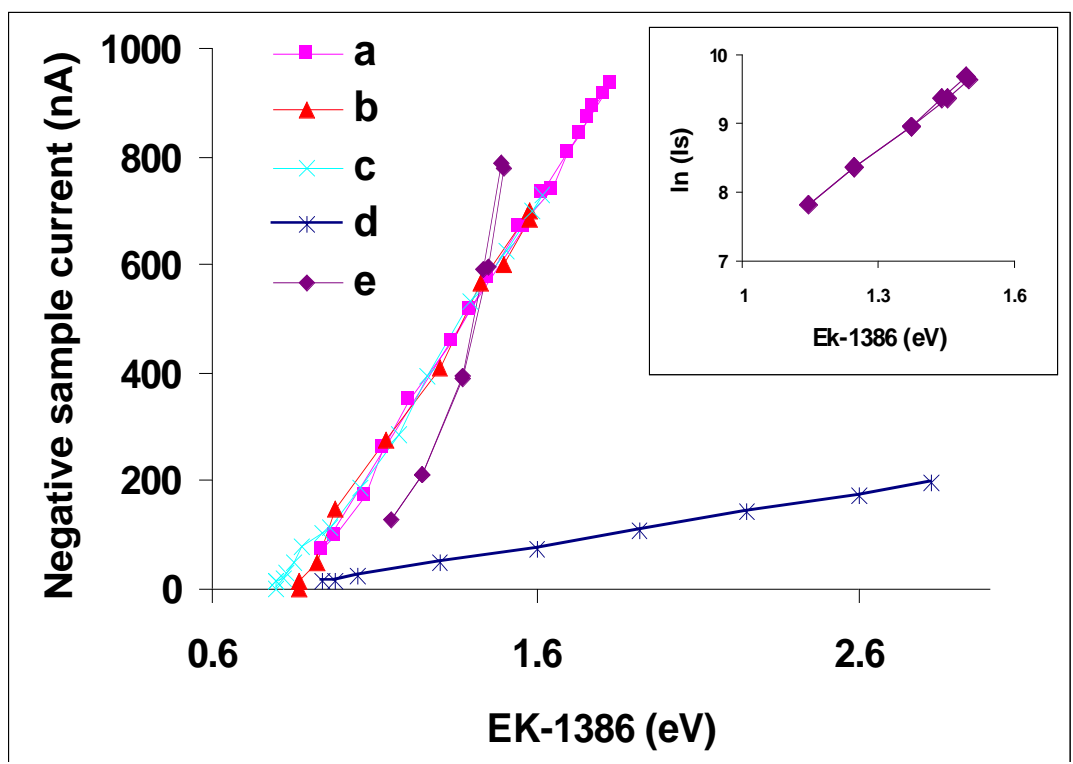

Fig. 2

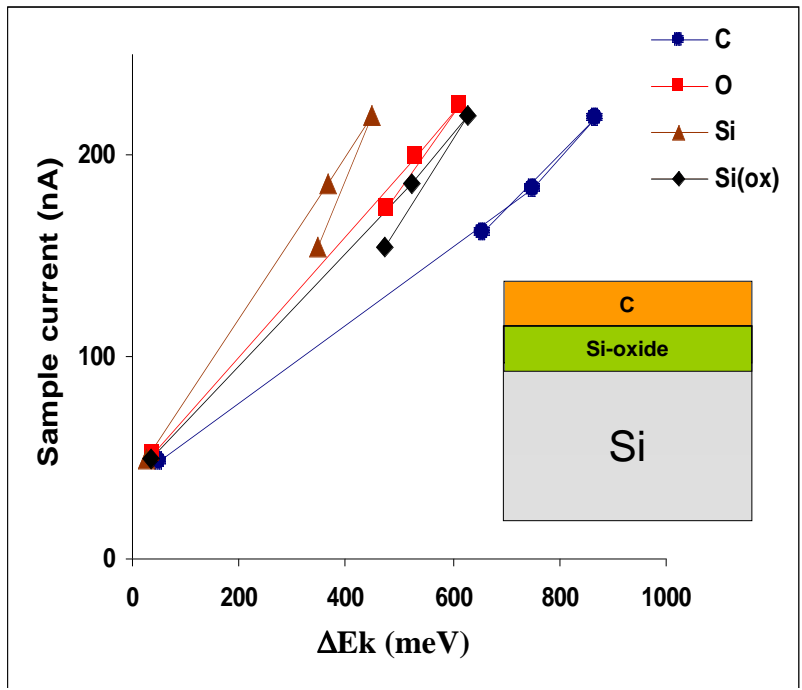

Fig. 3

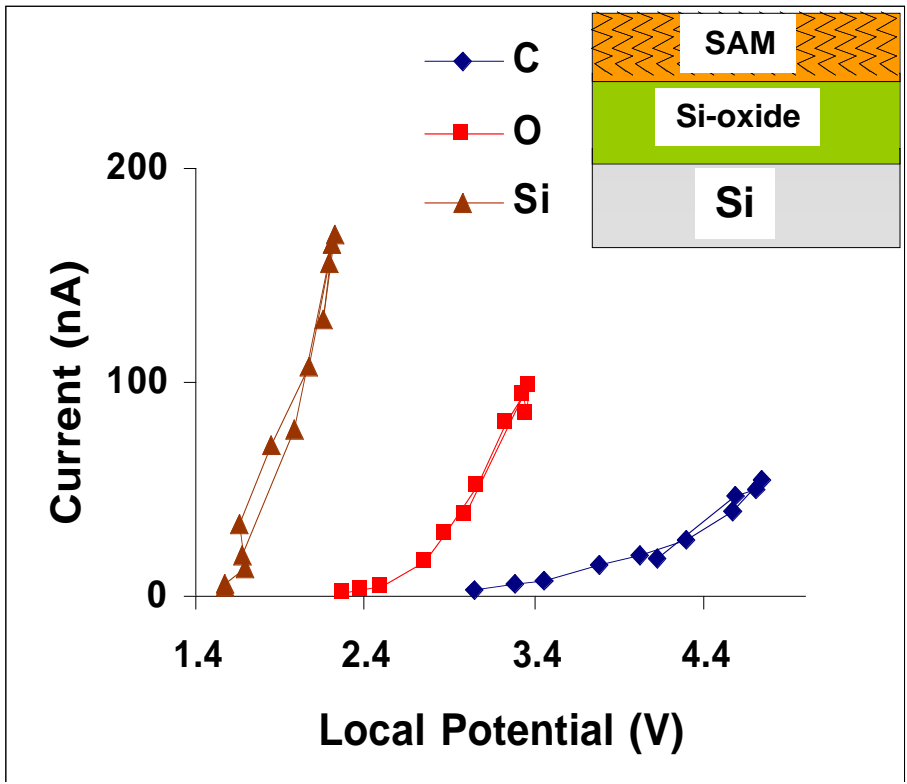

Fig. 4a

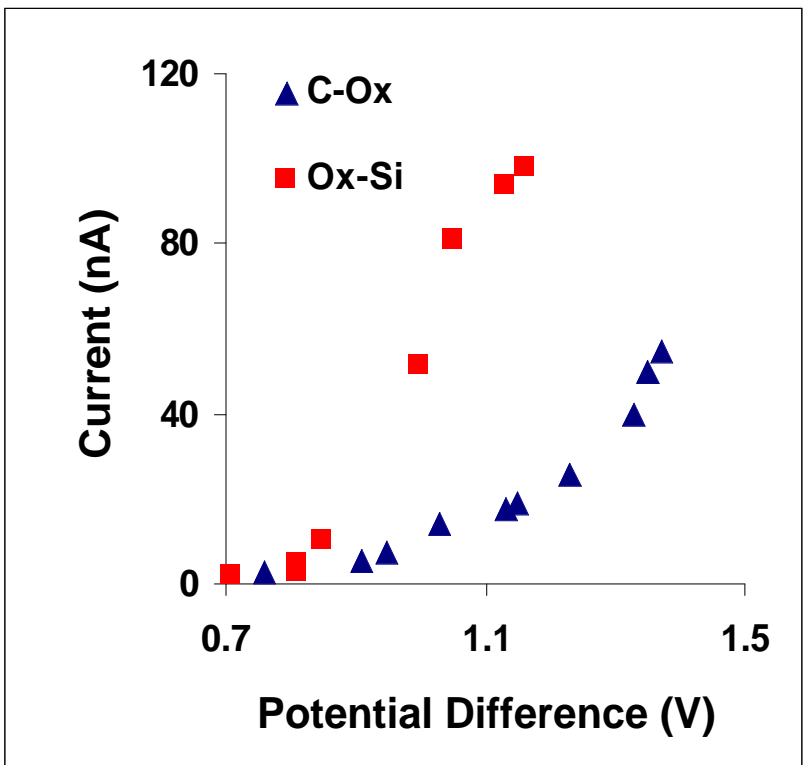

Fig. 4b